\documentclass[twocolumn,prl,aps,superscriptaddress]{revtex4}

\usepackage{amsmath,amssymb,latexsym,revsymb}
\usepackage[pdftex]{graphicx}
\pdfoutput=1

\def\duzomniejsze{<\kern-.7mm<}
\def\duzowieksze{>\kern-.7mm>}

\def\textbf#1{{\bf #1}}
\def\beq{\begin{equation}}
\def\eeq{\end{equation}}
\def\be{\begin{equation}}
\def\ee{\end{equation}}
\def\ben{\begin{eqnarray}}
\def\een{\end{eqnarray}}
\def\beqa{\begin{eqnarray}}
\def\eeqa{\end{eqnarray}}
\def\eea{\end{array}}
\def\bea{\begin{array}}
\newcommand{\bei}{\begin{itemize}}
\newcommand{\eei}{\end{itemize}}
\newcommand{\bee}{\begin{enumerate}}
\newcommand{\eee}{\end{enumerate}}

\newcommand{\tr}{\operatorname{Tr}}

\def\>{\rangle}
\def\<{\langle}

\def\rab{\rho_{AB}}
\newcommand{\ket}[1]{| #1 \rangle}

\def\12{{\textstyle \frac{1}{2}}}

\newcounter{protoline}
\newlength{\boxwidth}
\newlength{\bigboxwidth}

\newtheorem{proto_body}{Protocol}

\newenvironment{protocol}[1]{ 
  \setcounter{protoline}{0}
  \begin{minipage}{\boxwidth}
    
    \begin{proto_body}[ #1 ]
      \ \\[0mm] 
      \begin{description} }{\end{description}\end{proto_body}
\end{minipage}}
\newenvironment{protocol_cont}[0]{ 

\begin{minipage}{\boxwidth}
\addtocounter{proto_body}{-1}

\begin{proto_body}
\ \\ 
\begin{description} }{\end{description}\end{proto_body}
\end{minipage}}

\begin{document}

\title{State redistribution as merging: introducing the coherent relay}

\begin{abstract}

State redistribution allows one 
party to optimally send part of her state to another party.  Here
we show that this can be derived simply from two applications of
coherent state-merging. This provides a protocol whereby a middle party
acts as a relay station to help another party more efficiently transfer quantum states.  
This also gives a protocol for state splitting and the reverse
Shannon theorem (assisted or unassisted by side information), and 
allows one to use less classical communication for partial state-merging using a sub-protocol we call {\it ebit repackaging}.  
Thus state-merging 
generates the other primitives of quantum communication theory, reducing
the hierarchy between members of the first family of quantum protocols.
\end{abstract}

\author{Jonathan Oppenheim}
\affiliation{Department of Applied Mathematics and Theoretical Physics, University of Cambridge U.K.}

\maketitle

In \cite{devetakyard-redistribution,devetakyard-redistribution-longer} the problem of state redistribution was considered.
Namely, Alice and Bob share a quantum state, and Alice
wants to send part of her state to Bob by sending
only quantum states and using pre-shared entanglement. In such a situation
Alice can use the part of the state she doesn't send to Bob in order to send less than if she
didn't have access to this part.  The proof of this, and the resulting protocol were fairly
complicated~\cite{devetakyard-redistribution-longer}. Here, we show a simple and transparent 
protocol for state redistribution 
using state-merging~\cite{how-merge,how-merge2}.  This leads 
to another way to organise the family of protocols~\cite{fqsw} 
which form the basic building blocks of quantum communication theory.  It also provides a new protocol for several
other common tasks including a version of quantum
state merging using less classical communication in the case that part of the state remains at the sender's site. 

In state redistribution, Alice and Bob share $n$ copies of state $\rho_{ABC}$
with Alice holding onto $\rho_{AC}=\tr_B \rho_{ABC}$.  
One imagines a total 
pure state $\ket{\psi}_{ABCR}$
by introducing a reference system $R$.  
%
The task is for Alice to transfer $\rho_A$ to Bob while otherwise keeping the overall state 
$\ket{\psi}_{ABCR}^{\otimes n}$ virtually unchanged (in terms of fidelity).  The protocol is allowed to consume
(or produce) ebits i.e. shared entanglement in state $\ket{\phi^+}=(\ket{00}+\ket{11})/\sqrt{2}$.

If all that is available to Alice and Bob is a quantum channel, then
Alice can send her share $A$ to Bob using
$n I(A:R|B)/2$ qubits~\cite{devetakyard-redistribution} where the mutual information is defined as $I(C:R)=S(C)+S(R)-S(CR)$
and $I(A:R|B)=S(A|B)+S(R|B)-S(AR|B)$ with $S(A|B)$ the conditional entropy
defined as $S(AB)-S(B)$.  The optimal protocol also consumes ebits at a rate of 
$E= I(A:C)/2 - I(A:B)/2$.  If this
quantity is negative, that this amount of entanglement is produced.


We will show that redistribution uses {\it quantum state merging} as a basic primitive. In state merging 
Alice and Bob share $n$ copies of state $\rho_{AB}$, and 
Alice is able to optimally transfer her state to Bob using $n S(A|B)$ ebits
and $n I(A:R)$ bits of classical communication (the result $x$ of a random measurement she performed on her state).  

Now, to perform merging with a quantum channel instead of a classical one,
the sender is forced to send the classical measurement result $x$ using the quantum channel.  This gives
a coherent version of merging, sometimes called the Fully Quantum Slepian Wolf Theorem (FQSW)
or {\it merging mother}~\cite{devetak-fqsw,fqsw}.  One can derive this coherent-merging from the original merging protocol
by sending the classical communication using super-dense coding~\cite{BennettWiesner}.  This consumes $nI(A:R)/2$
ebits, and requires that $n I(A:R)/2$ qubits be sent. However, Alice could also make the measurement 
coherently i.e. perform a cnot operation from her state to an ancilla prepared in the $\ket{0}$ state 
and store the measurement result $x$ as the state
$\ket{x}\ket{x}$.  One half of this state can then be encoded and sent using super-dense coding.   
Since the measurement result $x$ is independent of
the final state after the protocol and is distributed uniformly, this generates
$nI(A:R)/2$ ebits (i.e. $\sum_x \ket{x}\ket{x}$).  Adding these generated ebits to the initial $nS(A|B)$, 
yields a total gain of $nI(A:B)/2$ ebits and a cost of $I(A:R)/2$ sent qubits. 
%
%
%

A direct protocol for achieving this rate\cite{fqsw} is for Alice to apply a random unitary $U$ to her state and
an ancilla of $nI(A:R)/2$ qubits initialised to $\ket{0}$, and then send this ancilla to Bob.  
Bob then performs a decoding unitary
$V$ on his system and the sent qubits.  As a result, Bob will possess $\rab$ and the two parties 
share $I(A:B)/2$ ebits.
 
A naive application of coherent-merging in the case when Alice also holds share $\rho_C$
would require n$I(A:RC)/2$ qubits to transfer $\rho_A$ to Bob since $\rho_C$ would be treated as the reference system 
to which correlations have to be maintained.  However, if Alice makes use of $\rho_C$ 
then redistribution only requires
$nI(A:R|B)/2$ qubits, a saving of $nI(A:C)/2$ qubits.

We now show that a less naive application of coherent-merging can be used as a primitive 
to perform state redistribution.
The essence of the idea is that Alice should not attempt to send all of $\rho_A$ to Bob, and in
particular should not send the pure state entanglement which exists between $A$ and $C$.  This pure state entanglement
can instead be extracted at a rate of $I(A:C)/2$, 
and replaced by ebits which were pre-shared between Alice and Bob, thus reducing
the number of qubits which have to be sent.  One shouldn't waste the quantum channel to transfer
ebits.  
 
\begin{figure}[htbp]
  \fbox{\includegraphics[width=8cm]{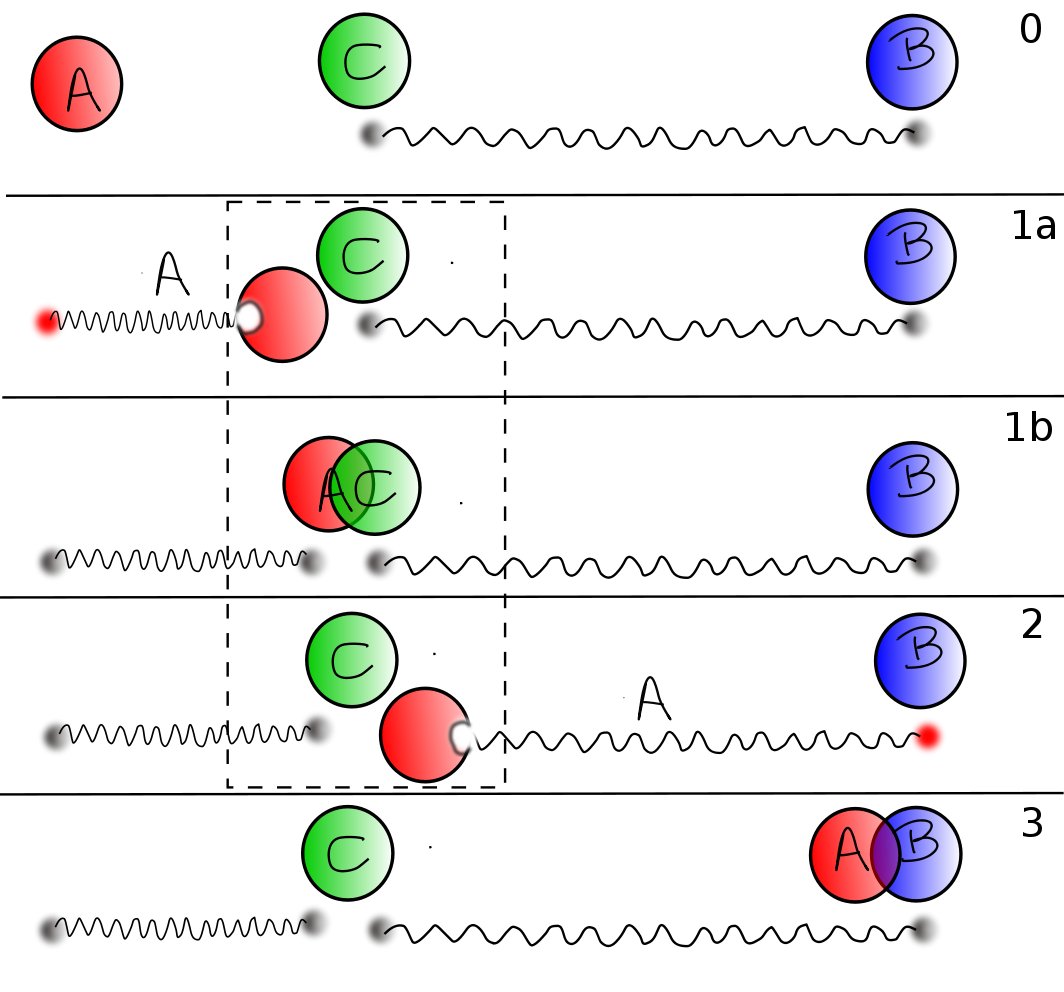}}
   \caption{The protocol.  Shares of state $\rho_{ABC}$ are represented by circles, while shared
entanglement is represented by wiggly lines.  The key difference between the situation depicted in
$1a$ and $2$ is that
$I(A:C)/2$ qubits (highlighted in red) have in effect been transferred from
Alice to Bob.  The steps refer to those in the protocol below and the sub-protocol of repackaging is
contained in the dashed square.}

  \label{fig:protocol}
\end{figure}

For the purpose of the protocol we will imagine that Alice is split into two
parties, one holds $\rho_A$ (we will give this party the name Alice) and 
Charlie who holds $\rho_C$.  As a result,
we will actually get a more general three party protocol with Charlie acting as a relay station,
minimising the number of qubits sent to Bob.

\vspace{.5\baselineskip}

\centerline{{\bf Protocol: redistribution from merging}}
\begin{itemize}
\setlength{\itemsep}{-\parskip}

\item[\bf 1:] Alice coherently merges $\rho_A$ to Charlie.  This extracts 
a rate of $I(A:C)/2$ ebits, and uses the quantum channel between
Alice and Charlie at a rate of $I(A:RB)/2$.  We can break this into two steps. 
{\bf 1a:} Alice applies the random unitary $U$ and sends to Charlie $nI(A:RB)/2$ qubits.  
{\bf 1b:} Charlie applies the decoding unitary $V$ on his state and the qubits from Alice, generating
$I(A:C)/2$ ebits between himself and Alice
 
\item [\bf 2:] Charlie sets aside the ebits that were generated from the previous step 
and replaces them with ones shared between him and Bob.  He then applies $V^\dagger$.
    
\item [\bf 3:] The effect of the previous step is no different from Alice having transferred $n I(A:C)/2$ qubits
to Bob.  Charlie then sends the remaining $nI(A:CR)/2-nI(A:C)/2$ qubits needed to transfer $\rho_A$ to
Bob.  This leaves him with $nI(A:B)/2$ qubits which are in fact ebits between himself and Bob.
\end{itemize}

The key element is that the qubits that Alice sent to Charlie are no different to the ones she kept behind,
except in the amount.  Thus, after step $2$, the situation is completely equivalent to Alice
having sent $nI(A:C)/2$ qubits.
Since a naive 
coherent-merging protocol requires $nI(A:RC)/2$ qubits to be sent from Alice to Bob, 
and the {\it ebit repackaging} performed in 
steps $1a-2$ are completely equivalent
to Alice having already sent  $nI(A:C)/2$ qubits to Bob, all that remains to be sent are 
$nI(A:R|B)= nI(A:RC)/2-nI(A:C)/2$ 
qubits.  Accounting for sent qubits $Q^{A\rightarrow C}$ between Alice and Charlie 
and $Q^{C\rightarrow B}$ between Charlie and Bob, as well as  
consumed ebits $E_{AC}$ and $E_{BC}$ we have the optimal rate pairs 
\begin{eqnarray*}
Q^{A\rightarrow C}= \12 I(A:RB)&, & E^{AC} = \12 I(A:C) \\
Q^{C\rightarrow B}= \12 I(A:R|B)&, & E^{CB} = \12 I(A:C) - \12 I(A:B) 
\end{eqnarray*}

Interestingly, in the sub-protocol of ebit repackaging 
(steps $1a-2$), $\rho_C$ is needed but is not changed, acting as a catalyst.  Ebit repackaging 
can be used in other protocols. If one performs repackaging before applying the random measurement
used in state merging (on what remains of $\rho_A$), 
then it reduces the amount of classical communication needed 
in the case when only $\rho_A$ is merged -- only $I(A:R|B)$ classical bits
are used, rather than $I(A:RC)$.   The case of redistribution when $\rho_B$ is null is called 
state-splitting (or the Fully Quantum
Reverse Shannon Theorem).
The time reverse of state-redistribution is a coherent version of the reverse Shannon theorem aided by side-information
at a relay station.  Repackaging gives a protocol for these tasks as well.

It was previously believed that state redistribution was a more general primitive which
could be used to construct the other building blocks of quantum communication theory, 
such as coherent-merging, which could then be used to construct the merging protocol 
and the so-called father protocol as well
as many others.  Here, we see that 
a number of primitives can generate the other building blocks of quantum Shannon theory --
we have shown how merging can be used
as a primitive to construct state redistribution and coherent-merging.  
Likewise, coherent-merging can generate redistribution and merging.

{\bf Acknowledgements.} I thank Igor Devetak for interesting discussions.
This work is supported by the Royal Society, and
EU projects SCALA and QAP.
\bibliographystyle{apsrev}

\end{document}